\documentclass{cjaa}                    % preprint, the final version
\usepackage{graphicx}                   %for PS/EPS graphics inclusion
\begin{document}
\title{The relation between X-ray spectral index and the Eddington ratio
in AGNs}
\author{Wei-Hao Bian}
\institute{$^{1}$Department of Physics, Nanjing Normal University,
Nanjing 210097, China
\\
$^{2}$National Astronomical Observatories, Chinese Academy of
Sciences, Beijing 100012, China}

\date{Received ...; accepted ...}
\titlerunning{}
\authorrunning{W. Bian}

\abstract{Using the H$\beta$ linewidth, we obtained the virial
central supermassive black hole masses and then the Eddington
ratios in a sample of broad-line AGNs and NLS1s observed by ASCA.
Combined with the data from ROSAT and Chandra observations, We
found a strong correlation between hard/soft X-ray photon index
and the Eddington ratio. Such a correlation can be understood by a
two-zone accretion flow model, in which zone is a thin disk and
the inner zone is an advection-dominated accretion flow (ADAF)
disk. The relation between X-ray photon index and the Eddington
ratio may account for NLS1s with not too steep X-ray photon index
founded by SDSS. If this relation is directly related to the
accretion disk, it may also exist in the accretion disk of
different scales (such as microquasar) 
          }
\maketitle

\section{Introduction}
With the reverberation mapping technology, there appears a rapid
progress on the mass estimate of the supermassive black hole (SBH)
in active galactic nuclei (AGNs) and then the Eddington ratio (the
ratio of the bolometric luminosity to the Eddington luminosity,
$L_{bol}/L_{Edd}$), which boosts our understanding of their
central structure and their evolution.

Steep soft X-ray photon index is a character of NLS1s compared
with the broad-line AGNs (Boller, Brandt \& Fink 1996). A popular
model of NLS1 is that they contain less massive black holes, but
have higher accretion rates radiating at close Eddington
luminosity. Within the Sloan Digital Sky Survey (SDSS) Early Data
Release (EDR) 150 NLS1s are founded (Williams, Pogge \& Mathur
2002) and the soft X-ray photon index of some NLS1s observed by
Chandra are found not too steep compared with that normally
observed in NLS1s (Williams, Mathur \& Pogge 2004).

Here we suggested that there exists a relation between X-ray
photon index and the Eddington ratio in AGNs including broad-line
AGNs and NLS1s. The Eddington ratio will lead to smaller soft
X-ray photo index in some NLS1s.

\section{Relation between X-ray photon index and the
Eddington ratio}

\subsection{Eddington ratio}

If we know the broad line region (BLR) size ($R_{BLR}$) and BLR
velocity ($v$), we can derived the black hole mass ($M$) using
Newton law, $M=V^{2}R_{BLR}G^{-1}$. The BLRs sizes can be derived
from the reverberation mapping method or the empirical
size-luminosity formula (Kaspi et al. 2000),
\begin{equation}
R_{\rm BLR}=32.9(\frac{\lambda L_{\lambda}(5100
\rm{\AA})}{10^{44}\rm erg~s^{-1}})^{0.7} ~~\rm{lt-days},
\end{equation}
where $\lambda L_{\lambda}(5100 \rm{\AA)}$ is the monochromatic
luminosity at 5100$\rm{\AA}$. Assuming the random BLRs orbits, the
BLRs velocity can be derived from the H$\beta$ linewidth
($v_{FWHM}$),
\begin{equation}
V=(\sqrt{3}/2) v_{\rm FWHM}.
\end{equation}

It is usually suggested there exists accretion disk system in the
center of AGNs. We used the Eddington ratio, $L_{bol}/L_{Edd}$, as
a direct measurement of the accretion rate. $L_{bol}$ is usually
calculated by $L_{bol}=9\lambda L_{\lambda}(5100 \AA)$(Kaspi et
al. 2000), where $L_{\lambda}(5100 \AA)$ is the monochromatic
luminosity at 5100$\rm{\AA}$.

\subsection{X-ray spectral index}

In general, the X-ray emission spectra of AGNs can be well fitted
by a power law ($f_{\nu}\propto \nu^{-\alpha}$ , where $\alpha$ is
the spectral index and $\Gamma = \alpha +1$ is the photon index).
The hard X-ray photon indices of AGNs observed by ASCA are
determined in the 2-10 keV (rest frame) for each source, excluding
the 5-7.5 keV (rest frame) which can contain significant photons
from an Fe K$\alpha$ line (confusing measurements of the hard
X-ray index). For soft X-ray spectral index, it is determined in
the 0.1-2.4 keV (rest frame). For very weak X-ray emission, the
photon index can be derived using the hardness ratio and other two
parameters: response function and column density (Williams et al.
2004).

\subsection{Samples}
\begin{figure}
\centerline{\includegraphics{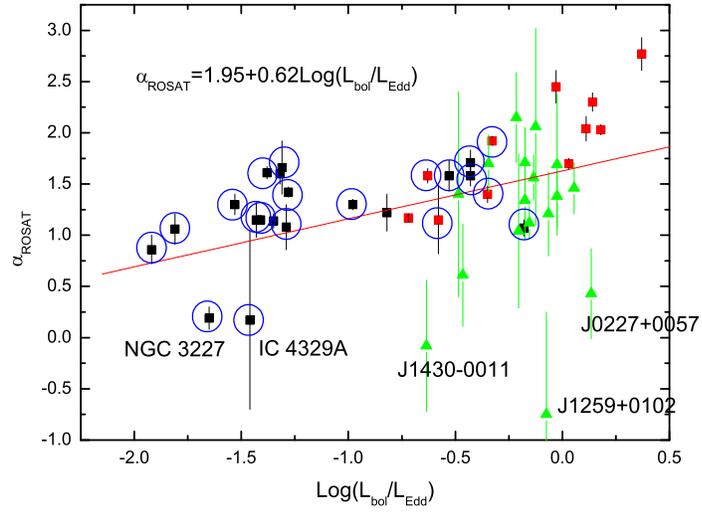}}

 \caption{\small The hard X-ray photo index versus the Eddington
ratio. The AGNs with the reverberation mapping masses are labelled
by blue circles. NLS1s are labelled by red squares. SDSS NLS1s
observed by Chandra are denoted by green triangles.}
% \vglue-0.9cm
 \end{figure}

\begin{figure}
\centerline{\includegraphics{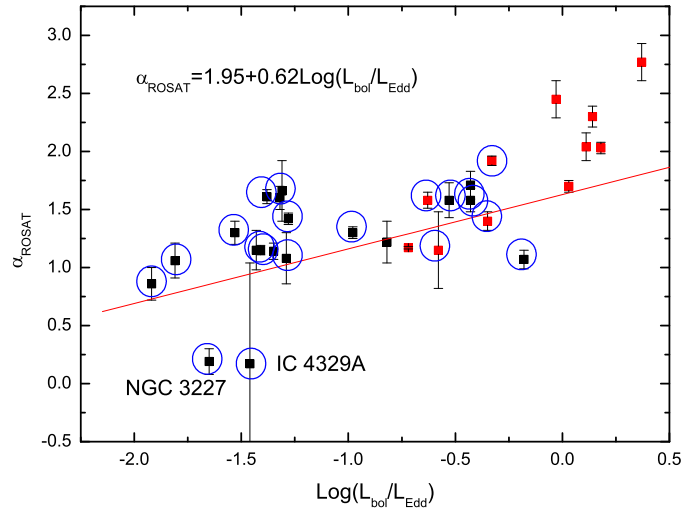}}

 \caption{\small The soft X-ray spectral index versus the Eddington
ratio. The denotation is the same as that in Fig. 1.}

 \end{figure}

\begin{figure}
\centerline{\includegraphics{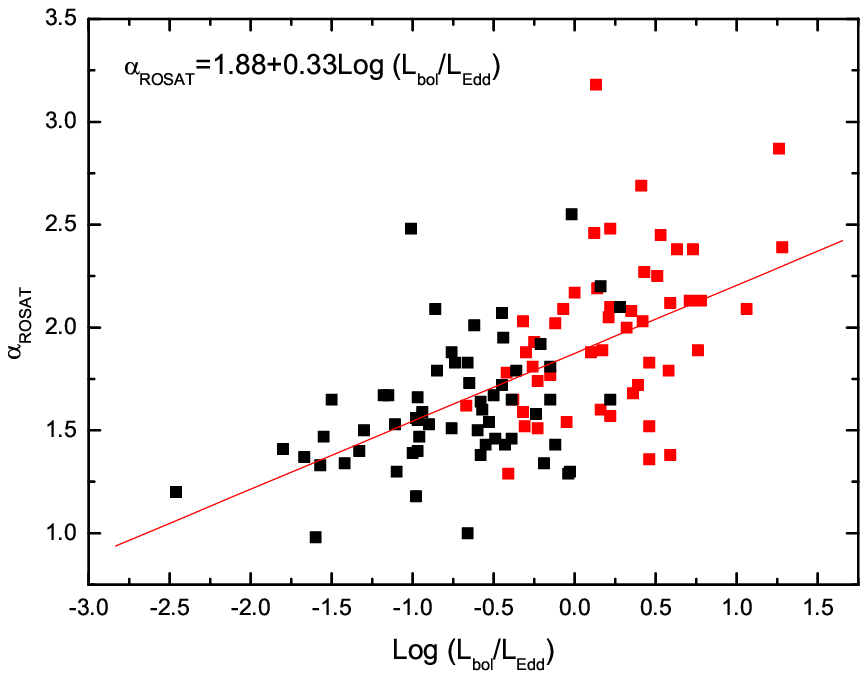}}

 \caption{\small
Soft X-ray spectral index versus the Eddington ratio for the
sample of Grupe et al. (2004)(Grupe 2004). NLS1s are labelled by
red squares.}

 \end{figure}

We have assembled a sample of broad-line AGNs and NLS1s to do a
research on the X-ray excess variability (Bian \& Zhao 2003). The
sample consists of 41 AGNs, in which there are 18 NLS1s and 23
broad-line AGNs.

For our sample (Bian \& Zhao 2003), the relation between the
hard/soft X-ray photon/ spectral index and the Eddington ratio is
showed in Fig. 1-2. A simple least-squares linear regression gives
$\alpha = 1.95+0.62Log(L_{bol}/L_{Edd})$ with a Pearson
correlation coefficient of $R=-0.74$ corresponding to a
probability of $P<10^{-4}$ that the correlation is caused by a
random factor. The best fit line is showed in Fig. 2 for all AGNs
in our sample.

Grupe et al. (2004) presented a complete sample of 110 soft X-ray
selected AGNs adopting the criterion of hardness ratio less than
zero and found about half of them are NLS1s. Fig. 3 showed the
soft X-ray spectral index versus the Eddington ratio for the
sample of Grupe et al. (2004) (Grupe 2004). A simple least-squares
linear regression gives $\alpha = 1.88+0.33Log(L_{bol}/L_{Edd})$
($R=0.59$, $P<10^{-4}$). The best fit line is also showed in Fig.
3.

Williams et al. (2003) presented a sample of 150 NLS1s found
within SDSS EDR, which is the largest sample of NLS1s. Williams et
al. (2004) recently present Chandra observations of 17
optically-selected, X-ray weak narrow-line Seyfert 1 (NLS1)
galaxies and exhibit a range of 1.1-3.4 for the 0.5-2 keV photon
indices, which extend to values far below what are normally
observed in NLS1s. From Fig. 2, we found that most of these 17
NLS1s follow the fit line. The calculated Eddington ratios showed
that their smaller photon indices are related to their smaller
Eddington ratios.

\section{Discussion}
From Fig. 2, we can find that some AGNs deviated much from the
trend. It is possibly due to the uncertainty of the estimate of
bolometric luminosity from optical luminosity. Different energy
mechanics and/or different accretion state would lead to different
energy distribution. It is simple to derive the bolometric
luminosity from the monochromatic luminosity at 5100 $\AA$. There
is some uncertainty in the estimate of black hole mass. For NLS1s,
the orbits of BLRs is not random, we may make some correction of
equation (2). Bian \& Zhao (2004) discussed the different methods
to estimate black hole masses in NLS1s.

The relation between X-ray spectral index and the Eddington ratio
can be understood in the frame of the accretion disk and the
corona. For a thin disk, the disk luminosity to irradiate the
corona increases as the Eddington rate increases. This can cause
the corona to cool efficiently owing to Compton cooling and cooler
corona producing few hard X-ray photons leads to large X-ray
spectral index. For the ADAF model of low accretion rate, the
optical depth increases and causes a correspondent increase in the
Compton $\gamma$-parameter when the Eddington ratio increases.
Larger optical depth will result in a harder and smoother X-ray
spectrum index, which is in conflict with our discussed relation.
A accretion disk consisting two zones (outer thin disk and inner
ADAF disk) is our preferable interpretation. The truncation radius
of the two zones decreases with increasing Eddington ratio. The
ADAF X-ray photon index becomes dramatically softer because the
radiation from the disk is Compton-scattered by the hot gas in
ADAF as the Eddington rate increases. This correlation possibly
gives a clue to the formation of corona (Wang et al. 2004).

It seemed that this relation is directly related to the accretion
disk, therefore it may also exist in the accretion disk of
different scales (such as accretion disk in microquasar).

\begin{acknowledgements}
This work has been supported by the NSFC (No. 10273007; No.
10273011) and NSF from Jiangsu Provincial Education Department
(No. 03KJB160060).
\end{acknowledgements}

\end{document}